\documentclass{emulateapj}

\shorttitle{The black hole in Cen~A with NACO}
\shortauthors{H\"aring-Neumayer et al.}

\begin{document}

\title{VLT Diffraction Limited Imaging and Spectroscopy in the NIR: Weighing the
  black hole in Centaurus~A with NACO \footnote{Based on observations collected
  at the European Southern Observatory, Paranal, Chile, ESO Program 72.B.0294A}}

\author{N. H\"aring-Neumayer\altaffilmark{1}, M. Cappellari\altaffilmark{2},
  H.-W. Rix\altaffilmark{1}, M. Hartung\altaffilmark{3},
  M.~A. Prieto\altaffilmark{1}, K. Meisenheimer\altaffilmark{1} and
  R. Lenzen\altaffilmark{1}} 

\altaffiltext{1}{Max-Planck Institute for Astronomy, K\"onigstuhl 17, 69117
  Heidelberg, Germany}
\altaffiltext{2}{Leiden Observatory, Postbus 9513, 2300 RA Leiden, The
  Netherlands} 
\altaffiltext{3}{European Southern Observatory, Casilla 19, Santiago, Chile}

\begin{abstract}
We present high spatial resolution near-infrared spectra and images of the
nucleus of Centaurus~A (NGC~5128) obtained with
NAOS-CONICA at the VLT. The adaptive optics corrected data have a spatial
resolution of 0$\farcs$06 (FWHM) in K- and 0$\farcs$11 in H-band, four times
higher than previous studies. The mean velocities and velocity dispersions of
the ionized gas ([FeII]) are mapped along four slit positions. The [FeII]
emission line width decreases from a central value of $\sigma\simeq320$ km s$^{-1}$ to
$\sigma\simeq120$ km s$^{-1}$ at $\rm{r}\simeq0\farcs5$, and exceeds the mean rotation
within this radial range. The observed gas motions suggest a kinematically hot
disk which is orbiting a central object and is oriented nearly perpendicular
to the nuclear jet. We model the central rotation and velocity dispersion
curves of the [FeII] gas orbiting in the combined potential of the stellar
mass and the (dominant) black hole. Our physically most plausible model, a
dynamically hot and geometrically thin gas disk, yields a black hole mass of
$\rm{M_{bh} = 6.1^{+0.6}_{-0.8} \times 10^{7}M_{\sun}}$. As the physical state of
the gas is not well understood, we also consider two limiting cases: first a
cold disk model, which completely neglects the velocity dispersion, but is in
line with many earlier gas disk models; it yields an $\rm{M_{bh}}$ estimate
that is almost two times lower. The other extreme case is to model a
spherical gas distribution in hydrostatic equilibrium through Jeans
equation. Compared to the hot disk model the best-fit black hole mass
increases by a factor of 1.5. This wide mass range spanned by the limiting
cases shows how important the gas physics is even for high resolution
data. Our overall best-fitting black hole mass is a factor of 2-4 lower than
previous measurements. A substantially lower $\rm{M_{bh}}$ estimate when using
higher resolution kinematics was also found for many other black hole mass
measurements as HST data became available. With our revised $\rm{M_{bh}}$
estimate, Cen~A's offset from the $\rm{M_{bh}}$-$\sigma$ relation is significantly reduced; it
falls above this relation by a factor of $\sim2$, which is close to the
intrinsic scatter of this relation. 
\end{abstract}


\section{INTRODUCTION}

Galaxy merging, the formation of stellar spheroids, nuclear star-formation and
the 
fueling of nuclear black holes appear to be all linked, forming a central
theme in building galaxies. Most galaxies where these processes are currently
acting, are so far away that the `sphere of influence' of the black hole
cannot be spatially resolved and that detailed studies of stellar populations
are impossible. Yet, our own Galactic Center alone tells us that galaxy
centers tend to become increasingly more interesting when observed at higher
and higher spatial resolution \citep[e.g.][]{schoedel02, genzel03}. When zooming
into the nucleus of Centaurus~A (NGC~5128), recent HST imaging (NICMOS) and
ISAAC 
observations have also revealed smaller and smaller `sub-systems'
\citep{schreier98, marconi00, marconi01}. Cen~A, the closest massive
  elliptical galaxy, the nearest 
recent merger, and one of the 
nearest galaxies with a significantly active nucleus provides a unique
laboratory to 
probe the interconnection between these phenomena on scales that are contained
in the central resolution element in any other object of its kind.\\
The intricate dust lane that hides the center of Cen~A has not allowed optical
high-resolution spectroscopy with 
HST. Infrared (IR) spectroscopy is thus increasingly important to open up the
regime of dust shrouded nuclei and get accurate black hole mass measurements
for these objects.\\
Cen~A is especially interesting since the black hole mass deduced by
\cite{marconi01} (M$_{\rm{bh}} = 2.0^{+3.0}_{-1.4}\times 10^{8}\rm{M}_{\odot}$)
lies a factor of ten above the $\rm{M_{bh}}$-$\sigma$ relation 
\citep{ferrarese00, gebhardt00}. This is the largest offset from the
relation measured to date. It is important to check this value,
since it can help us to understand the coevolution of black holes and their
surrounding bulges in more detail.\\
We have initiated a program to study the central parsec of Cen~A using
NAOS/CONICA \citep{rousset98, lenzen98} at the Very Large Telescope
(VLT).  
The adaptive-optics assisted
imager and spectrograph provides us with near infrared data from 1 to 5 $\mu$m at the
diffraction limit of a 8m class telescope. The resolution is thus nearly
fourfold that of HST in K-band.\\ 
The distance to Cen~A is still under discussion. In a comprehensive review
\cite{israel98} gives a value of 3.40$\pm$0.15 Mpc. \cite{tonry01} finds a
value of 4.2$\pm$0.3 Mpc from I-band surface brightness fluctuations, and
recently, \cite{rejkuba04} derived a distance of 3.84$\pm$0.35 Mpc from Mira
period-luminosity relation and the luminosity of the tip of the red giant
branch.  
Here, we assume a distance of D=3.5 Mpc to be consistent with the black hole
mass measurements of \cite{marconi01} and \cite{silge05} that also used that
value.\\
The paper is structured as follows: in Section 2 we present the observational
strategy and the data reduction. Section 3 describes the treatment of the
spectral data. Section 4 presents the dynamical modeling and the results for
the black hole mass for Centaurus~A, and Section 5 discusses the implications
of these results.

\section{OBSERVATIONS AND DATA REDUCTION}

\subsection{Adaptive Optics Observations}

Near infrared observations were performed in 2004 March 28 and 31 with
NAOS-CONICA (NACO) at the Yepun unit (UT4) of the very large telescope
(VLT). NACO consists of the high-resolution near-infrared imager and
spectrograph CONICA \citep{lenzen98} and the Nasmyth Adaptive Optics System
(NAOS) \citep{rousset98}. It provides
adaptive-optics corrected observations in the range of 1-5 $\mu$m with
14$\arcsec$ to 54$\arcsec$ fields of view and 13 to 54 mas pixel scales.\\
The data were taken in visitor mode and seeing during observations was in the
range 0$\farcs$3-0$\farcs$8 (as measured by the seeing monitor in V-band), with clear/photometric conditions. 

Not seen in the visible, the active nucleus at the center of Centaurus~A is an
unresolved source in K-band of 10.9 mag as detected by \cite{marconi00}.
There are no potential reference stars bright enough ($m_K\le$14 mag) for the
wavefront correction at a distance of $\le 30\arcsec$ to the 
nucleus, necessary for a good quality of correction at
the nucleus. Therefore, we directly guided on the nucleus itself using the
unique IR wavefront sensor implemented in NAOS. This strategy provides us the
best possible wavefront correction in the vicinity of the active galactic
nucleus (AGN). In fact we
reach the diffraction limit of the VLT in K-band of FWHM $0\farcs057$ and are
not far off in H-band with $0\farcs11$. 
During the observations the atmospheric conditions were stable and the
performance of the IR wavefront sensor (WFS) was steadily very good. For
observations in H-band we used the K-dichroic, i.e. all the nuclear
K-band light for wavefront correction. While observing in K-band itself the
only possibility to achieve a good performance of the WFS was to send 90\% of
the light to NAOS and only 10\% to CONICA (i.e. use the N90C10 dichroic). This
increases the exposure times by a factor of 10 and made it effectively
impossible to go for spectroscopy in K-band.

\subsection{K-Band Imaging}
\begin{figure}[ht]
\plotone{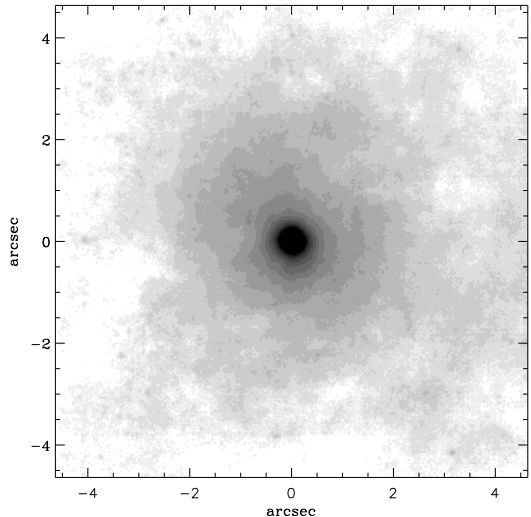}
\caption{K-band image taken with NACO. The nucleus is clearly visible as the
  unresolved source (FWHM=0$\farcs$06). North is up and East is to the
  left. \label{k-band}}  
\end{figure}

For the imaging we chose the strategy to jitter the field on five positions at
a separation of 4$\arcsec$ and take a sky image at a dark position in the
dust lane at a distance of $\sim$170$\arcsec$ South-East of the nucleus; this
cycle was repeated 4 times. The on-chip exposure time was 120~s, yielding a
total exposure time on the nucleus of 40~min. The resulting 
K-band image is shown in Figure \ref{k-band}.\\
The atmospheric conditions were stable and the seeing
was $0\farcs5$ at the start and $0\farcs8$ at the end of the observations.

\subsection{H-band spectroscopy}
\begin{figure}[ht]
\plotone{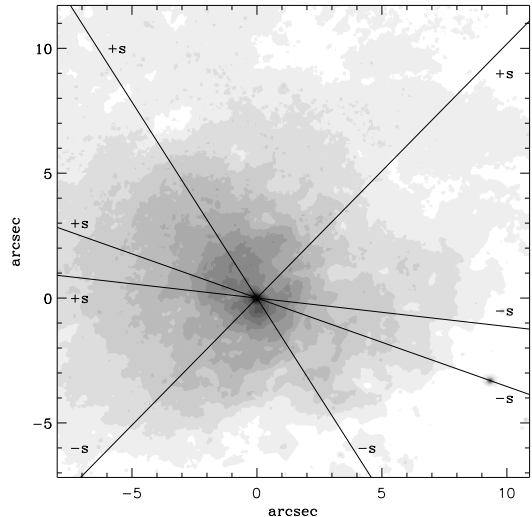}
\caption{Slit positions overlaid onto the H-band acquisition image. North is up
  and East is to the left. From North to South the slit positions are
  32$\fdg$5, 70$\fdg$5, 82$\fdg$5, and -44$\fdg$5. The sign of the position
  along the slit used in the rotation curves is indicated by ``+s''
  and ``-s''. Note the foreground star that is in the
  slit at P.A.=70.5$\degr$ in the lower right.\label{aqui}}  
\end{figure}

We took H-band spectra at 4 different position angles and chose three
similar to \cite{marconi01} (P.A.= -44$\fdg$5, 32$\fdg$5, and 82$\fdg$5) in
order to complement their findings with better 
spatial resolution. The fourth slit position (P.A.=70$\fdg$5) was chosen as to
observe the 
nucleus and a foreground star simultaneously to monitor the point spread
function (PSF) on a stellar
reference. See Figure \ref{aqui} for the positioning of the slits.\\ 
The observations were obtained with the $0\farcs086$ slit and a grating
with a wavelength range from $\lambda = 1.37\mu$m to $1.84\mu$m. The pixel scale was $0\farcs054$ leading to
a slight under-sampling but shorter exposure times. The dispersion was 7.0
$\AA$ pixel$^{-1}$, yielding a resolution of R=1500 in H.
At a given position angle, the observations went as follows: first, the
loop for adaptive optics correction was closed. When a good performance
of the WFS was reached, an acquisition image in H-band was taken. The
slit was centered on the prominent nuclear peak with an accuracy of
$\pm 0.2$pixel ($\pm0\farcs011$). The actual observations consisted of two
sequences of exposures at 4 positions along the slit. This was done in order
to perform sky subtraction and to avoid detector defects. The effective slit
length is $14\arcsec$. The exposure time per frame was 300~s, leading to a
total exposure time of 40~min per slit position angle.\\
Data reduction was performed using standard IRAF routines. The frames were
first bias corrected and flat-fielded with spectroscopic lamp-flats. Then
cosmic rays were rejected and the frames were wavelength calibrated and
corrected for distortions using spectroscopic arc lamps.\\
To get the final 2-D spectrum the frames were aligned. This is done by
shifting all frames to the same nuclear position.
Finally, the frames were sorted by quality of wavefront correction, using both
the width of the peak and the level of continuum flux. \\

\subsection{PSF-reconstruction}
A difficult but crucial part of adaptive optics
observations is the assessment of the point spread function (PSF). The PSF is
highly dependent on the quality of the wavefront correction, quantified
e.g. by the Strehl ratio. The Strehl ratio is set by the observable properties
of the reference object (flux, size, contrast) but 
also by the atmospheric conditions (seeing, coherence time);  it is
therefore changing with time and needs to be monitored throughout the
observations. In case of the K-band image a separate PSF reference star was
observed directly after the nucleus with the same WFS setup. This star
was chosen from the 2MASS point source catalogue \citep{cutri03} to match
Cen~A's nucleus as closely as possible: in angular proximity, magnitude and
color.   

Since the acquisition for NACO observations takes a non-negligible amount of
time, going back and forth to the PSF star is a tedious and time-consuming
task. On the other hand the measurement of a separate PSF-star at a given
difference in time and position on the sky does not guarantee to give a good
approximation on the actual on-source PSF. We therefore came to the conclusion
that it is most suitable to measure the PSF in the science frame on the
unresolved nucleus.
 
For emission line spectroscopy the strategy is to have the nucleus in the slit
at all position angles and measure the PSF on this unresolved point-source for
each frame. We tested this approach by choosing one position angle (P.A.$=70\degr$) such that
the nucleus plus a foreground star are simultaneously observed, and indeed, the
widths of both light profiles are similar (compare Table~\ref{Table1}).

\begin{figure}[ht]
\epsscale{0.8}
\plotone{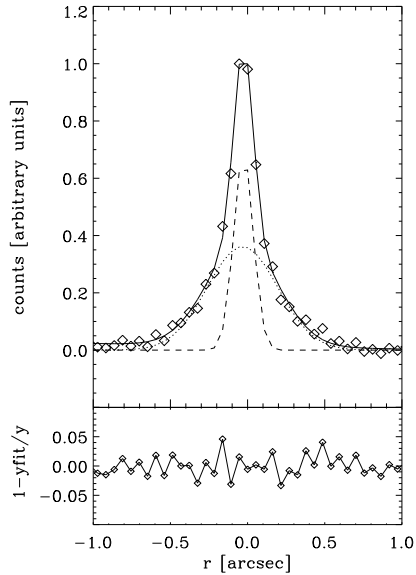}
\caption{Decomposition of the PSF for the slit position angle
  82$\fdg$5. The data are over plotted by the 2-component PSF-model (solid
  line), where the dashed line indicates the corrected peak and the dotted
  line represents the uncorrected seeing halo. The bottom panel shows the
  residual between model and data. \label{psf_decomp}}  
\end{figure}

\begin{figure}[ht]
\epsscale{1.0}
\plotone{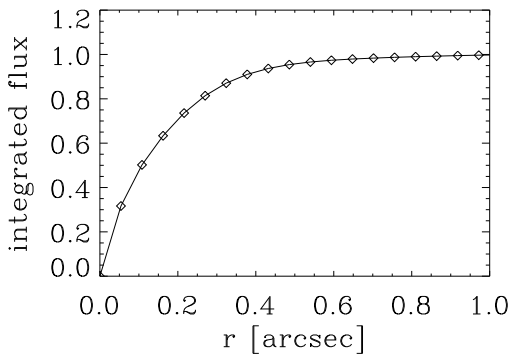}
\caption{Integrated flux over the model PSF is shown for the slit position
  angle 82$\fdg$5 (to be compared with Figure \ref{psf_decomp}). \label{psf_decomp_flux}} 
\end{figure}

We describe the normalized PSF empirically by a sum of two Gaussians; one narrow component
describing the
corrected/almost diffraction limited PSF core ($\sigma_{\rm{DL}}$) and one
broader component which we later attribute to the seeing halo ($\sigma_{\rm{s}}$): 
$$PSF(r) = \rm{ \frac{F}{2\pi \sigma^2_{DL}}e^{-r^2/2\sigma^2_{DL}} + 
 \frac{(1-F)}{2\pi \sigma^2_{s}} e^{-r^2/2\sigma^2_{s}}, }$$ 
where F is the ratio of the flux of the narrow component and the total flux of
 the PSF (F = $\rm{flux_{DL}/flux_{total}}$). The quantity F provides a rough  
approximation of the Strehl ratio (S) which gives the quality of an optical
system; S is defined as the observed peak flux divided by the 
theoretically expected peak flux of the Airy disk for the optical system
(S = $\rm{peak flux_{DL}/peak flux_{Airy}}$). 
For the following analysis it is sufficient to measure the quantity F, which
also gives an estimate of the quality of the adaptive optics correction.
\\
The width of the fitted broad component can be compared with the seeing 
that is given in the header for each frame, as measured by the seeing monitor
in V-band. One has to adjust the seeing estimates to the same wavelength, as
the resolution $\theta$ depends on wavelength $\lambda$, as
$\theta(\lambda)\sim \lambda^{-1/5}$. The individual components are given in 
Table \ref{Table1} and are compared to the header information (adapted to
H-band). The agreement between $\sigma_{\rm{s}}$ and $\sigma_{\rm{DIM}}$ is
satisfactory in all cases. Notice also the good agreement between the width
of the nucleus and the star at P.A.=70$\fdg$5. The adaptive optics correction
is optimised for the position of the nucleus and is worse at the position of
the star due to anisoplanetism.\\
Moreover it is obvious that the width of the narrow component depends on the
seeing conditions.\\
Figure \ref{psf_decomp_flux} shows the integrated flux over the two-component
model PSF shown in Figure \ref{psf_decomp}. Note that 50\% of the flux lie
within a radius of 0$\farcs$1 and 90\% of the flux within 0$\farcs$3.\\
In the PSF model we did not account for the undersampling of the observed PSF,
since the sampling problem is negligible compared to the general uncertainty
in adaptive optics observations.

\begin{table}[ht]
\begin{center}
\caption{Reconstruction of the PSF\label{Table1}}
\vspace{0.2cm}
\begin{tabular}{l c c c c c c }
\tableline\tableline
Object & P.A. & $\rm{FWHM_{DL}}$ & $\rm{FWHM_{s}}$ & $\rm{FWHM_{DIM(H)}}$ & F \\
\tableline
Nuc & 32$\fdg$5  & 0$\farcs$11  & 0$\farcs$37  & 0$\farcs$37 & 0.23 \\

Nuc & -44$\fdg$5 & 0$\farcs$15  & 0$\farcs$48  & 0$\farcs$48 & 0.18 \\

Nuc & 82$\fdg$5  & 0$\farcs$15  & 0$\farcs$53  &0$\farcs$49 &  0.15 \\ 

Nuc & 70$\fdg$5  & 0$\farcs$11  & 0$\farcs$34  & 0$\farcs$35 & 0.22 \\

Star & 70$\fdg$5 & 0$\farcs$12  & 0$\farcs$36  & 0$\farcs$35 & 0.31 \\

\tableline\tableline
\end{tabular}
\end{center}
\end{table}

\section{RESULTS}

\subsection{Nuclear spectrum}
\begin{figure}
\plotone{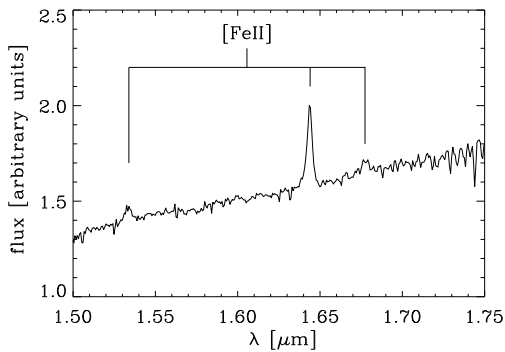}
\caption{Nuclear spectrum in H-band extracted from an aperture of $0 \farcs 054
\times 0\farcs 086$ at a slit position of 32$\fdg$5. The three [FeII]
emission lines are indicated  at the rest-frame wavelengths:
$\lambda_1 1.534\rm{\mu m}$, $\lambda_2 1.644\rm{\mu m}$, and $\lambda_3 1.677
\rm{\mu m}$. \label{nuc_spec}} 
\end{figure}

A nuclear H-band spectrum centered on the continuum peak (Fig. \ref{aqui}) and
extracted from a $0\farcs 054 \times 0\farcs 086$ aperture is presented in
Figure \ref{nuc_spec}. The spectrum exhibits a power-law continuum with
three [FeII] lines: $\lambda_1 1.534\rm{\mu m}$, $\lambda_2 1.644\rm{\mu m}$, and
$\lambda_3 1.677 \rm{\mu m}$. We use the strongest line ($\lambda_2 1.644\rm{\mu m}$) for our kinematical studies. 

\begin{figure}[ht]
\plotone{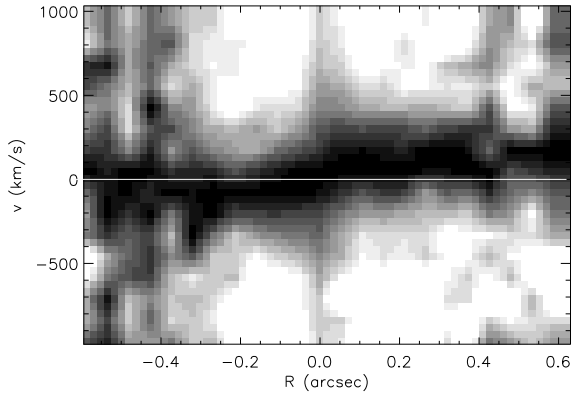}
\caption{Observed [FeII] $\lambda_2 1.644\rm{\mu m}$ line on the NACO H-band
  spectrum at a slit position of -44$\fdg$5. To make the gas velocity curve more
  visible, each column of the 
  spectrum was divided by the maximum value of the emission line in that
  column. The spectrum was then resampled on a 3 times finer grid by means of
  bilinear interpolation. The line indicates the systemic velocity of
  532 km s$^{-1}$, which is shifted to zero for this plot.\label{FeII_emission}}  
\end{figure}

Line-of-sight velocities, FWHMs, and surface brightnesses along each slit were
obtained by fitting single Gaussians to the [FeII] emission line in each row
of the continuum-subtracted two-dimensional spectra. The corresponding slit
positions are shown in Figure \ref{aqui} and listed in Table \ref{Table1}.

\subsection{Gas Kinematics}

\begin{figure*}
\plotone{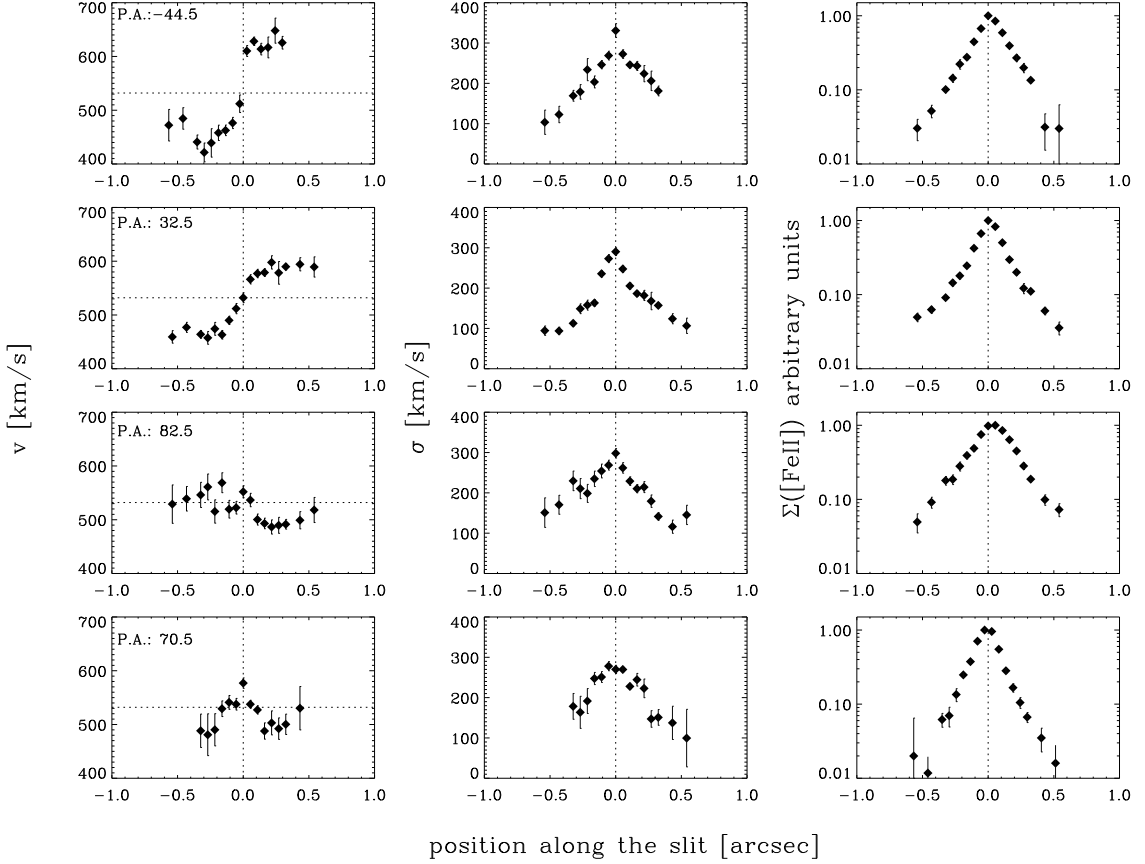}
\caption{Rotation curves (left panel), velocity dispersion profiles (middle
  panel) and emission line surface brightness profiles (right panel) along
  the 4 slit positions. For the position angles -44$\fdg$5 and 70$\fdg$5 some
  of the points shown in the surface brightness profile are missing for the v
  and $\sigma$ profiles. \label{rot}} 
\end{figure*}

We measure the gas kinematics on the ionised [FeII] line  $\lambda 1.644\rm{\mu m}$ out
to around 
$\pm0\farcs6$. Single Gaussians provide a good fit to the emission lines and are
used to measure the position and width of the line. The fit is performed in
IDL\footnote{See http://www.rsinc.com.} using a non-linear least squares fit to the line and the errors are the
1-$\sigma$ error estimates of the fit parameters.
The center of the continuum peak is taken as a reference for the systemic
velocity. We find a value of v$_{\rm{sys}}$=532 km s$^{-1}$; in good agreement with the
value v$_{\rm{sys}}$=532$\pm$5km s$^{-1}$ measured by \cite{marconi01} from
their gas kinematical data.\\ 
For the central 0$\farcs$3 only the four highest quality frames are considered
in order to make use of the full resolution. The quality of the frames,
i.e. their correction quality, is estimated on the basis of the width and the peak
value of their continuum peak. Outside 0$\farcs$3, when the lineflux drops, all
frames are taken into account and three pixels are binned to enhance the
signal-to-noise ratio. \\ 
The rotation of the gas can be directly seen in Figure~\ref{FeII_emission},
where the line intensity was normalised by the peak intensity in each
column. The rotation curves, velocity dispersion curves and emission-line
surface brightness profiles are shown in Figure \ref{rot} for all slit
positions.

\begin{figure}
\epsscale{1.0}
\plotone{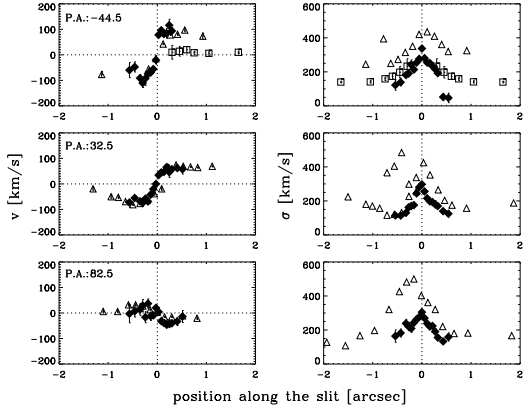}
\caption{Comparison between the spatial resolution of the NACO data (filled
  diamonds), the ISAAC data (open triangles) \citep{marconi01}, and the GNIRS
  data (open squares) \citep{silge05}. The NACO and ISAAC data correspond to
  gas kinematics while the GNIRS data to stellar kinematics. In the left panel the rotational
  velocity is shown and in the right the velocity dispersion.\label{compare}} 
\end{figure}

The velocity dispersion is directly measured as the width of the lines. We
corrected for instrumental broadening of 65 km s$^{-1}$, measured from the
skylines.\\
The excellent spatial resolution of the NACO data is demonstrated in Figure
\ref{compare} where we compare the NACO data points to the kinematical data published
by \cite{marconi01} and \cite{silge05}. The NACO velocity dispersions
are often considerably smaller than the ISAAC velocity dispersions measured
at the same location. The information given in \cite{marconi01} is not
sufficient to find the reason for this discrepancy and we believe that our
data points are correct.

\subsection{The Emission-Line Surface Brightness \label{sec_surface_brightness}}
The intrinsic surface brightness distribution of emission lines $\Sigma$ is an
important ingredient in the model computations because it is the weight in the
averaging of the observed quantities. It would be ideal to have an emission
line image with higher spatial resolution than the spectra. Unfortunately, the
resolution of the HST [FeII] narrow band image \citep{marconi00} is not good
enough to mimic the NACO data and therefore our approach
is to match the emission-line fluxes as observed in the spectra.  
We extracted the emission line surface brightness from [FeII] directly from
the spectra along the four slit positions (see Figure \ref{rot}, right
panels).\\
In order to get a parametrisation for the intrinsic surface brightness, we
test different functional forms and convolve them with the PSF of the
observations. The two-dimensional gas distribution is assumed to be a circular
disk. We find that the intrinsic surface brightness of a disk is well fitted
by a double exponential profile $$\rm{
I(r) = I_0 e^{-r/r_0} + I_1 e^{-r/r_1}, }$$ where $\rm{r_0}$ and $\rm{r_1}$
are the scale radii, and $\rm{I_0}$ and $\rm{I_1}$ are the scale factors for
the two components. This parametrisation was also used in \cite{marconi01}.
Since the observed emission line surface brightness along the slit at different
position angles depends on the inclination of the circular gas disk (with respect to the observer)
as well as on the PSF we fit the intrinsic emission-line fluxes for a median
disk inclination of $45\degr$ given the observational setup and
conditions. The projected major axis of the [FeII] disk is at P.A.=$32\fdg5$
\citep{marconi00} and gets the highest weight in the fit, since the surface
brightness along the major axis stays unchanged when the disk is inclined. 
The parameters that after PSF convolution best fit the
surface brightness distribution along the different 
slit positions are %I$_0$=6.5, I$_1$=0.12, 
I$_1$/I$_0$=0.018, r$_0$=0$\farcs$02, and
r$_1$=0$\farcs$22. This parametrisation for the disk surface brightness is
used in the dynamical models at all inclination angles. Compared to the fit
values in \cite{marconi01} our scale radii are smaller due to higher spatial
resolution. 
Given the shape of the
[FeII] emission from the HST narrowband image \citep{marconi00}, the axis
ratio of a possible gas disk is $\sim$1:2 which corresponds to an inclination
angle of 60$\degr$, given a circular disk structure. 
However, as found before by \cite{marconi01} there is a mismatch
between the photometric major axis of the [FeII] gas disk and its kinematical
major axis (line-of-nodes), which they find to lie between $-10\degr > \zeta >
-14\degr$. 

\section{DYNAMICAL MODEL}

In this section we outline the models describing the motions of the ionised
gas in the combined gravitational potential of the (putative) central black
hole and the stars. This modeling also requires a careful discussion of the
gas geometry, as well as the physical state of the ionised gas, whose measured
velocity dispersion far exceeds the expected thermal broadening.

\subsection{The Stellar Mass Model}\label{stellar_mass}
The first step in the dynamical modeling is to estimate the stellar
contribution to the central potential
from the stellar surface density. For the deprojection of the observed surface
brightness distribution into the stellar luminosity density we applied the 
Multi-Gaussian expansion (MGE)
method \citep{monnet92, emsellem94}. Specifically, we obtained an MGE
fit to a composite set of 2-dimensional K-band images using the method and
software of \cite{cappellari02}. 
The MGE fit was performed using the NACO K-band image
($0''<r<7''$), the NICMOS F222M image \citep{schreier98} ($2''<r<16''$), and
the 2MASS Large galaxy 
atlas (LGA) K-band
image \citep{jarrett03} for larger radii ($8''<r<365''$). \\
The sky-subtraction was performed relative to the 2MASS LGA image, which is
taken to be the sky-clean reference. The flux calibration of the NACO and
NICMOS data is also referenced to the 2MASS image.

To get a parametrisation of the NACO K-band PSF we fit a full MGE model to the
composite image of NGC~5128 (without giving the code a prior PSF estimate). We
then selected the central four Gaussian components, that clearly described the
unresolved nucleus, as the PSF. Its parametrisation is $\rm{PSF(R)=
\sum_{i=1}^{N}G_i \, exp[-R^2/
  (2\sigma_{i}^{*2})]/(2\pi\sigma_{i}^{*2})}$, and the numerical values of the
relative weights $G_{i}$ (normalised such that $\sum_{i=1}^{N}G_i =1 $) and of
the dispersions $\sigma_{i}^{*}$ are given in Table \ref{mge_psf_table}. 
This PSF was then
fixed in the MGE fit to the composite image, where we neglected PSF
convolution of the NICMOS and 2MASS images since these are only used at radii
$r>2''$ and the NACO image has a much higher resolution. \\  

\begin{table}[ht]
\begin{center}
\caption{MGE PSF parametrisation\label{mge_psf_table}}
\vspace{0.2cm}
\begin{tabular}{l c c c}
i & $G_{i}$ & $\sigma_{i}^{*}$ & FWHM$_{i}^{*}$ \\
\tableline\tableline
1..... & 0.0004 & 0$\farcs$005 & 0$\farcs$012 \\ 
2..... & 0.1590 & 0$\farcs$031 & 0$\farcs$074 \\   
3..... & 0.3034 & 0$\farcs$062 & 0$\farcs$145 \\
4..... & 0.5372 & 0$\farcs$143 & 0$\farcs$336 \\
\tableline
\tableline\tableline
\end{tabular}
\end{center}
\end{table}

\begin{figure}
\epsscale{1.0}
\plotone{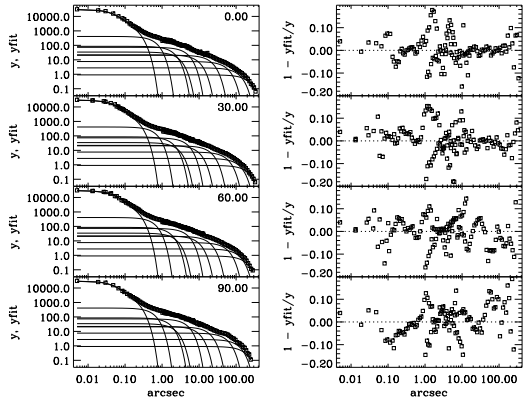}
\caption{Comparison between the NACO $(0''<R<7'')$, NICMOS $(2''<R<16'')$, and
  2MASS LGA $(8''<R<365'')$ K-band surface brightness (open squares) of Centaurus~A and the
  corresponding best-fit MGE model (solid line) as a function of radius
  R. On the left the MGE fit is shown for inclination angles between $0\degr$
  and $90\degr$, while the right panels show the corresponding residuals.\label{mge_fit}} 
\end{figure}

Figure \ref{mge_fit} shows a comparison between the observed photometry and
the MGE model along four different position angles in the galaxy, while Table \ref{mge_fit_table}
gives the corresponding numerical values of the analytically deconvolved MGE
parametrisation of the galaxy surface brightness $$
\Sigma(x',y') = \sum_{i=1}^{N}
\frac{L_i}{2\pi\sigma_i^2 q'_i}
\exp\left[-\frac{1}{2\sigma_i^2} \left(x'^2+\frac{y'^2}{q'^2_i}\right)\right],$$

where $(x^{'},y^{'})$ are the coordinates on the plane of the sky and N
is the number of the adopted Gaussian components, having total luminosity
$L_i$ , dispersion $\sigma_{i}$, and observed axial ratio 0.8$\leq q_{i}^{'}\leq$1.0.
In Table \ref{mge_fit_table} the central unresolved component (which we
attribute to the AGN) is removed. We only account for the stellar light
distribution in the Multi-Gaussian expansion model.

\begin{table}[ht]
\begin{center}
\caption{MGE parametrisation of the deconvolved K-band surface brightness of
  Cen~A\label{mge_fit_table}} 
\vspace{0.2cm}
\begin{tabular}{l r@{.}l r@{$\farcs$}l  r@{.}l}
\multicolumn{1}{c}{i} & \multicolumn{2}{c} {L$_i$ [$\times 10^9 L_{\odot,K}$]}
& \multicolumn{2}{c} {$\sigma_{i}$} & \multicolumn{2}{c} {q$^{'}_{i}$} \\ 
\tableline\tableline
1..... &  0&113 & 0&45 &  0&882\\ 
2..... &  0&297 & 1&48 & 0&812\\   
3..... &  1&13 & 3&27 & 1&000 \\
4..... &  1&25 & 6&68 & 1&000\\
5..... &  5&07 & 14&3 &  0&879\\ 
6..... & 15&5 & 38&0 &  1&000\\   
7..... & 19&6  & 78&6 & 1&000\\
8..... & 23&7  & 130&9 &  0&807\\   
9..... &  5&24 & 365&8 & 0&800 \\\tableline
\tableline\tableline
\end{tabular}
\tablenotetext{}{The total luminosity L$_i$, the dispersion $\sigma_{i}^{*}$,
  and the axial ratio q$^{'}_{i}$ for the nine Gaussian components that are
  needed to fit the surface brightness of Cen~A in the Multi-Gaussian
  expansion model. We overlap three K-band images (NACO $(0''<R<7'')$, NICMOS $(2''<R<16'')$, and
  2MASS LGA $(8''<R<365'')$) to model the surface brightness from 0'' to 365''
  (see Fig.~\ref{mge_fit}). The central component seen in Fig.~\ref{mge_fit}
  is omitted here.}
%\tablecomments{.}
\end{center}
\end{table}

Using this radial profile of K-band (volume) emissivity, we
fix the stellar mass-to-light ratio by constructing an isotropic spherical
Jeans model and matching it over the radial range $ 3''< r <10''$
to the stellar kinematics ($\sqrt{v_*^2+\sigma_*^2}$) published recently by
\cite{silge05}. The innermost $3''$ are excluded from the fit to minimise
the influence of the black hole. We find a best-fitting central mass-to-light ratio $\rm{M/L_K = (0.72 \pm 0.04)}$
  $\rm{M_{\odot}/L_{\odot}}$ in agreement with the best-fitting model of
\cite{silge05}. This agreement of the mass-to-light ratio from Schwarzschild
and Jeans modeling is supported by the recent work of \cite{cappellari2005}.\\

\subsection{Geometry and Kinematics of the [FeII] gas}
The central [FeII] velocity curves (Figure \ref{rot}) suggest that we see gas
rotating in a flattened geometry around a central mass concentration, presumably dominated
by a black hole.
The velocity 
dispersion in the very center is quite high ($\sim$300 km s$^{-1}$ in the central
pixel), well in excess of the mean rotation seen. In part this high dispersion
may be attributed to rapid but spatially unresolved rotation.\\
As the underlying physical origin of the observed high gas dispersion is not
clear a priori, we consider at least three geometric models:
\begin{enumerate}
\vspace{-0.2cm}
\item the gas lies in a geometrically thin, kinematically cold disk and is in
  Keplerian motion around the black hole,\\ 
\vspace{-0.6cm}
\item the gas forms a geometrically thin, kinematically hot disk that is in
  radial hydrostatic equilibrium, \\
\vspace{-0.6cm}
\item the gas lies in a spherical distribution of
collisionless cloudlets and can be modeled through Jeans equation. \\
\vspace{-0.6cm}
\end{enumerate}
Note that cases 1. and 3. are extreme physical assumptions and we consider
them as limiting cases. 
In all three cases the gas is moving in the combined potential of the
surrounding stars and of the central black hole; the self gravity of the gas
is negligible (given the estimated mass of
the ionised gas is $\sim 10^3 M_{\odot}$ \citep{marconi01} and the mass of the
innermost Gaussian representing the stars is already $\sim 8 \times 10^7 M_{\odot}$
(see Table \ref{mge_fit_table})). The underlying stellar
mass distribution is the same in all models, with M/L
fixed by the stellar kinematics at larger radii (see previous section).\\
The mismatch between the kinematical and photometric major axis of the gas
disk, found before by \cite{marconi01} introduces a degeneracy between the
inclination angle ($i$) of the gas disk and the position of its line-of-nodes
($\zeta$). We fear that our long-slit data will not be sufficient to constrain
both $i$ and $\zeta$. 
We therefore make use of the inclination angle of the radio jet which sets the
lower limit of the disk inclination in the standard picture of an orthogonal
disk-jet geometry. \cite{tingay98} derived a value of $50\degr < i <
80\degr$. Given this prior information, we set up our thin disk models.

\subsection{Model 1: Thin cold disk model}
\begin{figure*}
\plotone{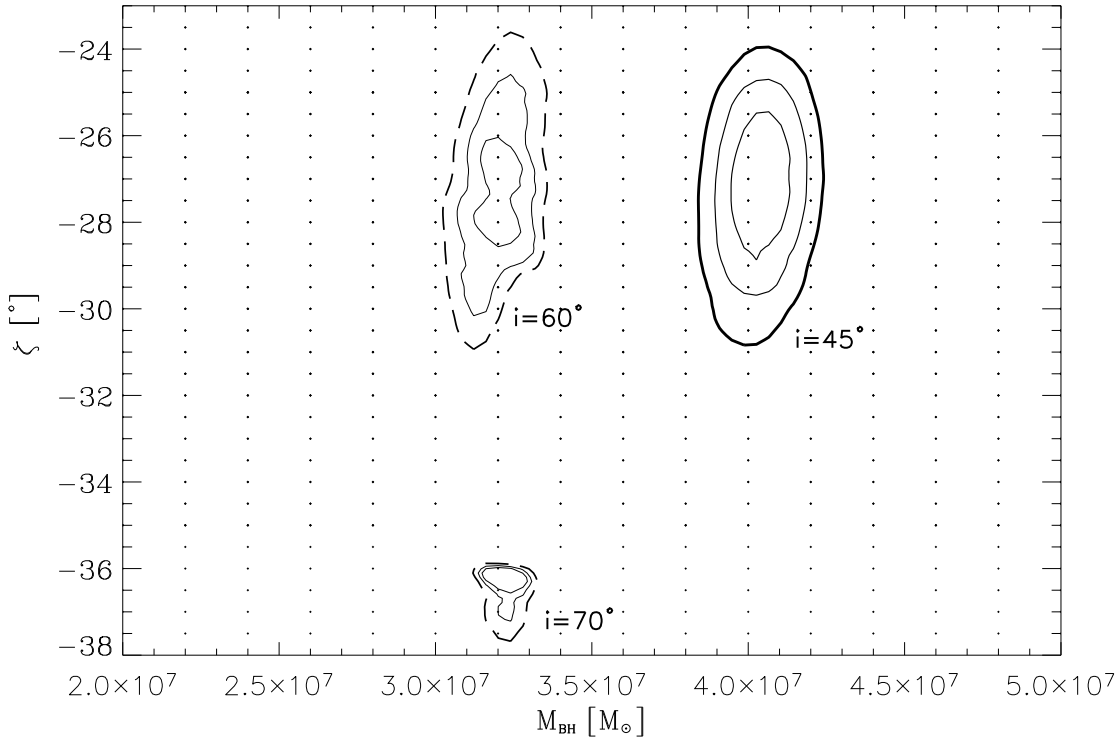}
\caption{Fitting a cold gas disk: the $\chi^2$ contours are shown for the two
  parameter fit of black hole mass vs. projected major axis $\zeta$ for the
  three inclination angles. The best cold model is at an inclination angle of
  $45\degr$ and the corresponding contours are plotted in solid lines. 
  The best-fitting values are $\zeta=-27\degr \pm 4\degr$ and 
  M$_{\rm{bh}} = (4.0 \pm 0.2)\times 10^{7}M_{\odot}$. The three
  contours give the formal 68.3\%, 95.4\%, and 99.73\% (thick) confidence
  levels.\label{contours_cold}} 
\end{figure*}

We follow the widely used approach \citep[e.g.][]{macchetto97} which assumes the
gas to lie in a thin 
disk around the black hole, moving on circular orbits; its observed
velocity dispersion is assumed to be solely due to rotation. We constructed a
model using the IDL software developed in 
\cite{cappellari1459}. This modeling can deal with multiple component PSFs,
different PSFs for 
different data-subsets, and a general gas surface
brightness distribution. Pixel binning and slit effects are taken into account
to generate a two-dimensional model spectrum with the same pixel scale as the
observations. Like in the data analysis, the rotational velocity and
velocity dispersion is determined by fitting simple Gaussians to each row of
the model spectrum. We make no velocity offset correction
\citep{vdmarel97, maciejewski01, barth01}, since the slit width 
($0\farcs086$) is comparable to the FWHM of the PSF.\\
The predicted velocity and velocity dispersion profiles of this model depend
on the intrinsic 
surface brightness distribution of the emission lines, the PSF, plus the
following parameters:
\begin{enumerate}
\vspace{-0.2cm}
\item the inclination, $i$, of the gas disk ($i=0\degr$ is face-on, 90$\degr$ edge-on)\\
\vspace{-0.6cm}
\item the angle between the projected major axis of the disk (line of nodes), $\zeta$, and the
slit positions\\
\vspace{-0.6cm}
\item the black hole mass, M$_{\rm{bh}}$; the stellar mass profile is fixed.\\
\vspace{-0.6cm}
\end{enumerate}
We describe the ``intrinsic velocity dispersion'' of the gas in the model by a
double exponential parametrisation of the form $$ \sigma_{R} = \sigma_{0}
e^{-r/r_0} (1+ \epsilon e^{-r/r_1}), $$ 
without using it in the
dynamics. The parameters $\sigma_{0}$, $\epsilon$, $r_0$, and $r_1$ are fixed through comparison of the
observed dispersion profile to a flux-weighted
convolution of the intrinsic velocity dispersion profile with the PSF and the
size of the aperture.  The
parameters that give a good fit to the velocity dispersion at all slit
positions in the cold disk model are,
$\sigma_{0}$=140 km s$^{-1}$ , $ \epsilon \simeq 0.25$,
$r_0=0\farcs9$, and $r_1=0\farcs2$. 
We do not know the source for this high velocity dispersion and 
simply ignore it in the kinematically cold disk model.
To reduce the number of free parameters in the model we evaluate the rotation
curves for three fixed inclinations ($45\degr$, $60\degr$, and
$70\degr$)
and fixed the intrinsic surface brightness density
of the gas disk beforehand (cf. Section \ref{sec_surface_brightness}). \\
In Table \ref{fit_values_cold} we give the best-fitting black
hole masses for each inclination angle individually; we also give the
resulting $\chi^2$ values to get the overall best-fitting model. The $1\sigma$,
$2\sigma$, and $3\sigma$
contours are shown in Figure \ref{contours_cold}
for the two degrees of freedom, black hole mass and line-of-nodes. The three
sets of contours correspond to the relevant inclination angles of the gas disk
(i=$45\degr$,$60\degr$, and $70\degr$)
where the best-fitting inclination angle, i=$45\degr$, is indicated by a solid
line.\\

\begin{table}[ht]
\begin{center}
\caption{Best Fit Values - Cold Disk Model\label{fit_values_cold}}
\vspace{0.2cm}
\begin{tabular}{c c c c}
i & M$_{\rm{bh}}$ [$M_{\odot}$] & $\zeta$ & $\chi^2_{\rm{tot}}$\\
\tableline\tableline
45$\degr$ & $( 4.0 \pm 0.2 ) \times 10^{7}$ & $-27\degr  \pm 4\degr$ & 46.1 \\
60$\degr$ & $( 3.2 \pm 0.2 ) \times 10^{7}$ & $-28\degr  \pm 4\degr$ & 47.1 \\
70$\degr$ & $( 3.2 \pm 0.1 ) \times 10^{7}$ & $-36\degr  \pm 2\degr$ & 47.1 \\
\tableline
\tableline\tableline
\end{tabular}
\end{center}
\end{table}

\begin{figure*}[ht]
\plotone{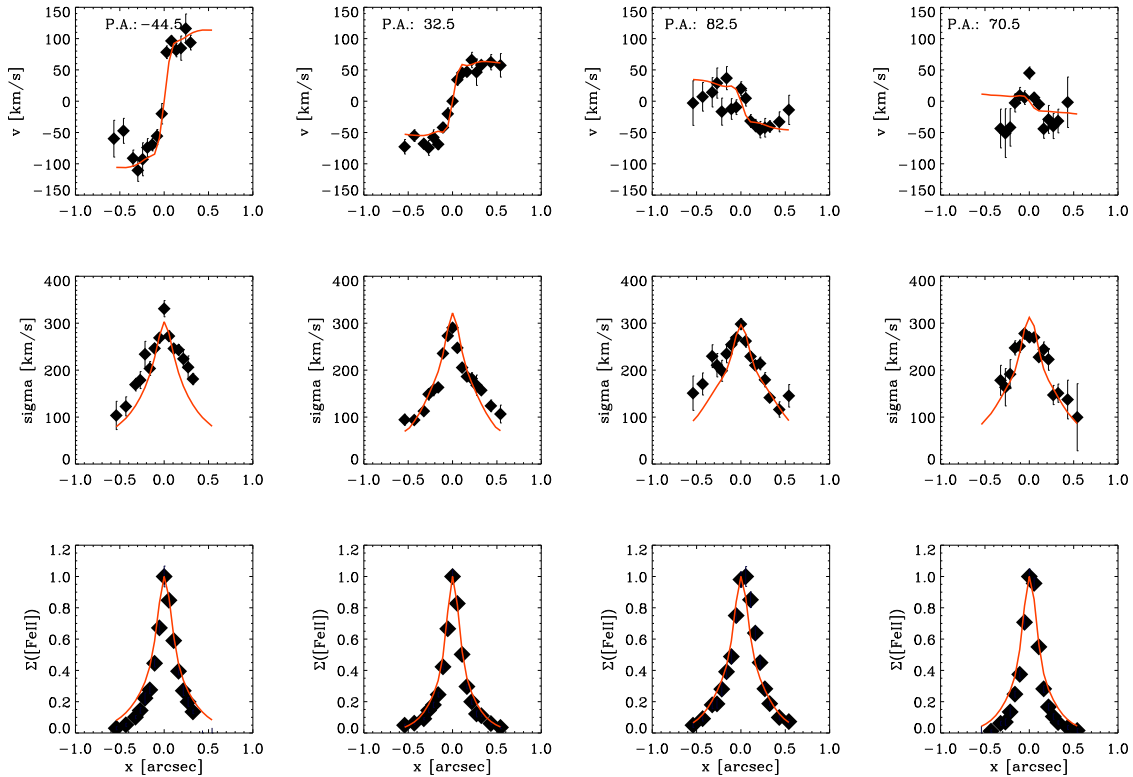}
\caption{The observational data is shown in comparison to the best fitting
  cold disk model at 45$\degr$ inclination angle. The best-fitting parameters
  are $\zeta=-27\degr$ and 
  M$_{\rm{bh}} = 4.0\times 10^{7}M_{\odot}$. The velocity dispersion as well
  as the surface brightness are parametrised by a double exponential function
  which were fixed beforehand and not fitted here. \label{best_fit_cold}} 
\end{figure*}

The overall best-fitting cold gas model is obtained at an inclination
angle of $45\degr$. Figure \ref{best_fit_cold} shows the observational data in
comparison to the model. The parametrisation of the surface brightness, shown
in the third row, is fixed beforehand and not included in the fit here
(compare Section \ref{sec_surface_brightness}). 
The parameters that lead to this fit are
$\zeta=-27\degr$ and M$_{\rm{bh}} = 4.0\times 10^{7}M_{\odot}$. \\
However, the minimum $\chi^2$ values are comparable for all three cases, i.e. their
differences are smaller than the 1$\sigma$ level, and we
therefore cannot constrain the inclination angle of the gas disk with our
long-slit data. 
We conclude that the black hole mass for a dynamically cold, geometrically
thin disk model is M$_{\rm{bh}} = (4.0^{+0.2}_{-1.0}) \times
10^{7}M_{\odot}$.

\subsection{Model 2: Thin hot disk model}

\begin{figure*}
\plotone{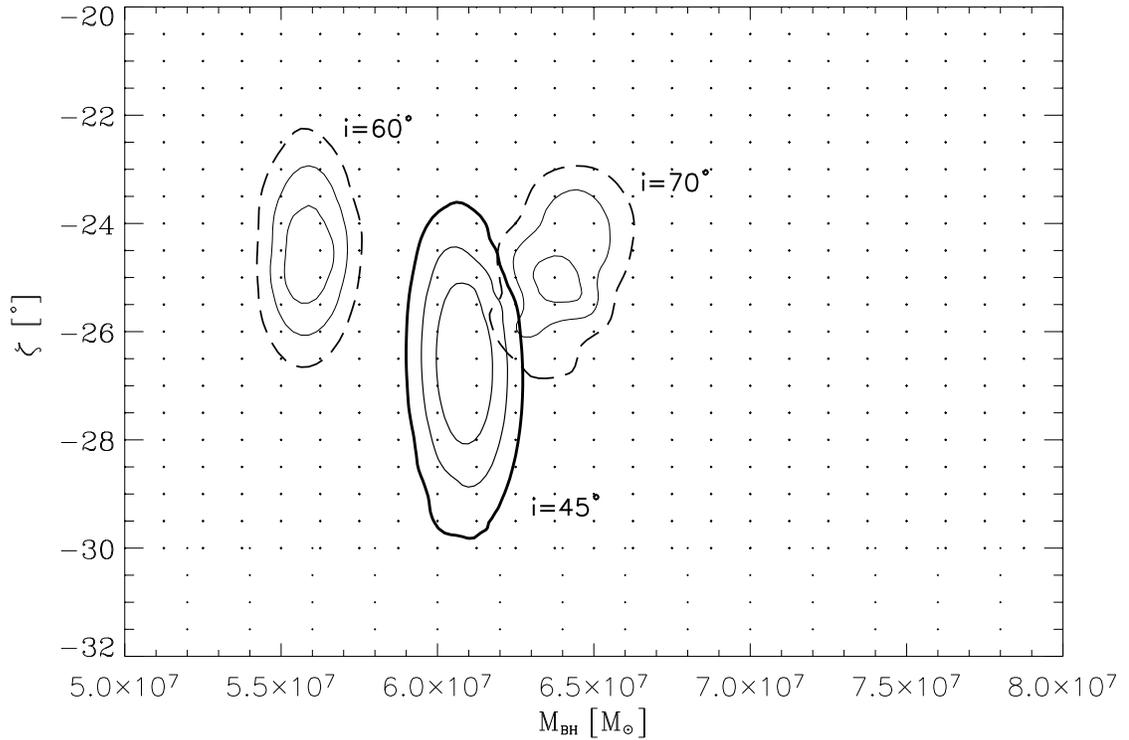}
\caption{Fitting the 'hot' gas disk: the $\chi^2$ contours are shown for the
  two parameter fit of black hole mass vs. projected major axis $\zeta$ for the three inclination
  angles. The best hot model is at an inclination angle 
  of $45\degr$ and the corresponding contours are plotted in solid lines.
  The best fitting values are $\zeta=-27\degr \pm 3\degr$ and 
  M$_{\rm{bh}} = (6.1 \pm 0.3)\times 10^{7}M_{\odot}$. The three
  contours give the formal 68.3\%, 95.4\%, and 99.73\% (thick) confidence
  levels.\label{contours_hot}} 
\end{figure*}

This model is identical to the previous one except that we now interpret the
``intrinsic velocity dispersion'' as a gas pressure component. As mentioned
above, we do not know the physical origin of this high velocity dispersion of
the gas, nevertheless, we include it in the dynamical analysis. \\
Any pressure support will require less rotation for dynamical equilibrium at a
given black hole mass. In other words, the black hole mass needs to be larger
to cause the same mean rotational velocity. The classic approach to account
for velocity dispersion is to apply an asymmetric drift correction (e.g 
\cite{barth01}), but the approximate equations are only applicable if $\sigma/v$ is
small. However, in the case of Cen~A, $\sigma/v$ even exceeds unity and we
therefore chose another approach: we assume the gas disk to be geometrically
flat but with an isotropic pressure; on this basis we construct an
axisymmetric Jeans model in hydrostatic equilibrium to model the observations.\\  
We assume that the [FeII] surface brightness of the gas disk $\Sigma_g(x,y)$
reflects the tracer gas density $\rho_g(x,y)$, and that the gas moves in the
joint potential $\Phi$ of the stars and the central black hole.
The Jeans equation for this situation reads \citep[][Eq.4-64a]{BT87}:
$$ \frac{R}{\rho_g} \frac{\partial(\rho_g \sigma_R^2)}{\partial R} + R
\frac{\partial\Phi}{\partial R} = \overline{v_{\Phi}}^2 ,$$\\
where R is the projected radius, and $\sigma_R$ and $\overline{v_{\Phi}}$ are
the radial velocity dispersion and the azimuthal velocity of the gas,
respectively, and both are functions of R.\\
We again parametrise the gas dispersion profile by a double exponential function
of the form: $$ \sigma_{R} =\sigma_{0} e^{-r/r_0} (1+ \epsilon e^{-r/r_1}),$$
and we fit for the best set of parameters ($\sigma_{0}$=140 km s$^{-1}$,
$\epsilon \simeq 0.25$, 
$r_0=0\farcs9$, and $r_1=0\farcs2$) to get the intrinsic dispersion profile.\\
The assumption that the gas disk is infinitesimally thin does most probably not
reflect the real physical properties but was chosen to eliminate line-of-sight
integrations through a 3-dimensional gas distribution.\\
Table \ref{fit_values_hot} summarises the best fit parameters and the
corresponding 
$\chi^2$ values for the different inclination angles of the ``hot'' gas disk
model. The $1\sigma$,
$2\sigma$, and $3\sigma$ contours are plotted in Figure \ref{contours_hot}.\\ 

\begin{figure*}
\plotone{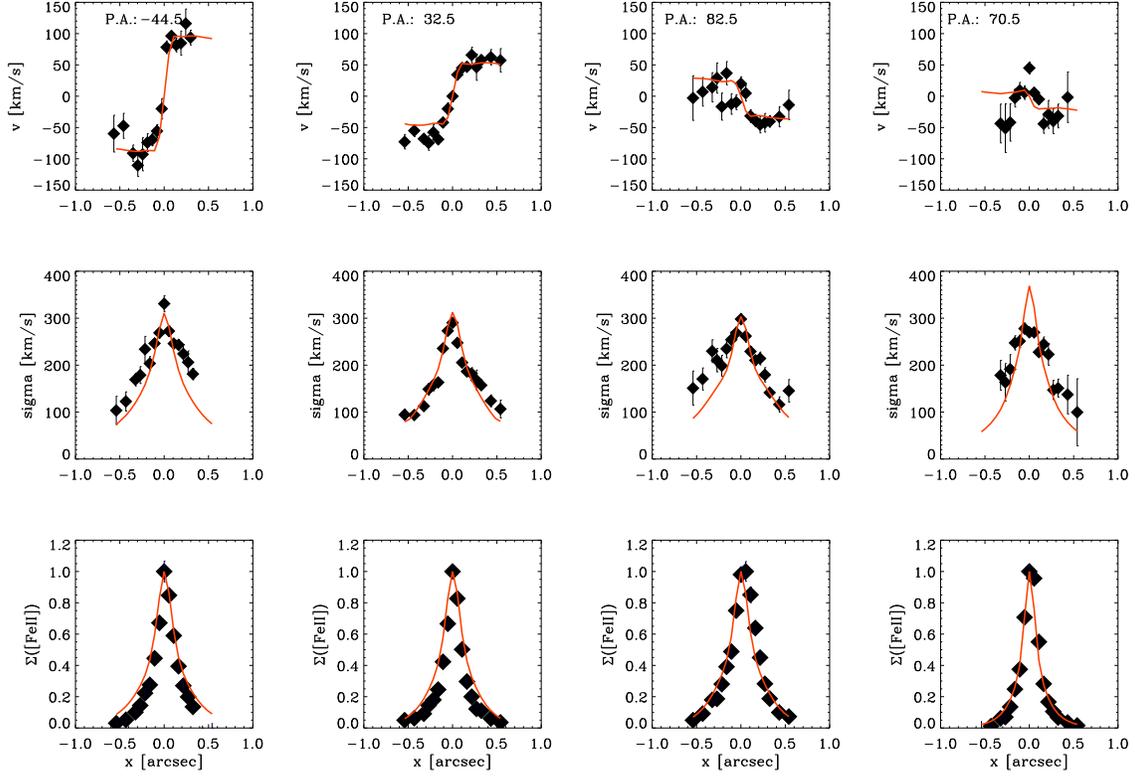}
\caption{The best-fitting
  hot disk model with parameters $\zeta=-27 \degr$ and
  M$_{\rm{bh}} = 6.1\times 10^{7}M_{\odot}$. In this model, $v$ and $\sigma$
  are interwined through Jeans equation. Note that despite the higher
  black hole mass (compared to the cold disk model), the model curves drop
  beyond $0\farcs3$ and match the observed data better. The model of the surface
  brightness is fixed beforehand and not fitted here. \label{best_fit_hot}} 
\end{figure*}

\begin{table}[ht]
\begin{center}
\caption{Best Fit Values - Hot Disk Model\label{fit_values_hot}}
\vspace{0.2cm}
\begin{tabular}{c c c c}
i & M$_{\rm{bh}}$ [$M_{\odot}$] & $\zeta$ & $\chi^2_{\rm{tot}}$ \\
\tableline\tableline
45$\degr$ & $( 6.1 \pm 0.3 ) \times 10^{7}$ & $-27\degr \pm 3\degr$ & 52.1 \\
60$\degr$ & $( 5.6 \pm 0.3 ) \times 10^{7}$ & $-25\degr \pm 2\degr$ & 54.0 \\
70$\degr$ & $( 6.4 \pm 0.3 ) \times 10^{7}$ & $-25\degr \pm 2\degr$ & 53.7 \\
\tableline
\tableline\tableline
\end{tabular}
\end{center}
\end{table}
The best-fitting hot disk model favors a black hole mass of M$_{\rm{bh}} =
(6.1 \pm 0.3) \times 10^{7}M_{\odot}$ and $\zeta=-27\degr \pm 3\degr$, at an inclination angle of
$45 \degr$. It is shown in Figure \ref{best_fit_hot} for comparison. 
The increased total $\chi^2$ value (compared to the cold disk model) can be
explained by the reduced number of degrees of freedom, since in the hot disk
model the velocity dispersion and the rotational velocity are coupled through
Jeans equation. 
However, the difference between the two fits is small ($< 2\sigma$
confidence level)
and the large velocity dispersion
lead us to consider this model a better description of the physical properties
at the center of Centaurus~A.\\

\subsection{Model 3: Spherical Jeans model}
\begin{figure}
\plotone{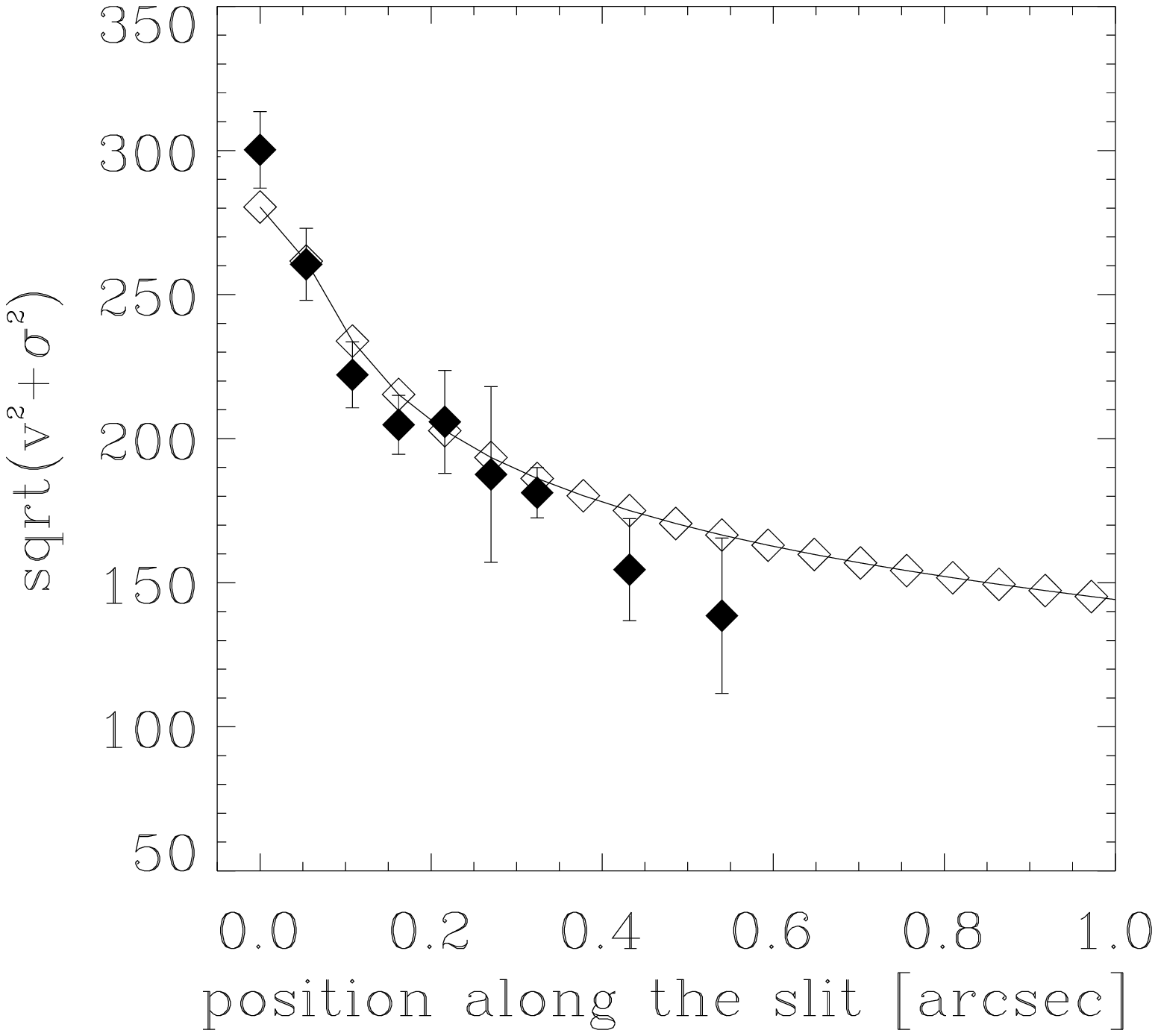}
\caption{Comparison between the measured rms velocity (filled diamonds)
  and the spherical Jeans model (open diamonds) for the gas. The best fitting black hole mass
  is M$_{\rm{bh}} = (1.0 \pm 0.5)\times 10^{8}M_{\odot}$. \label{spherical_jeans_gas}} 
\end{figure}

In this final modeling approach, we account for the fact that the high gas
velocity dispersion is not consistent with the assumption of a thin
disk. Alternatively, we assume the gas spherically symmetric
distributed 
in individual clouds that move ballistically (i.e. t$_{\rm{coll}}$$>>$t$_{\rm{dyn}}$). Given the observational facts
(rotation curves along certain slit positions, gas disk observed by
\cite{marconi00}), this model is not very likely to describe the physical
properties of the central region in Cen~A. 
However, it gives an upper limit on the black hole mass, compared with disk
models at non-extreme inclinations, and is implemented fairly easily. 
We construct a spherical Jeans model where we assume the
following:
\begin{enumerate}
\vspace{-0.2cm}
\item the gas density is given by its emissivity\\
\vspace{-0.6cm}
\item the gas cloud distribution is spherical \\
\vspace{-0.6cm}
\item the gas clouds move in the potential given by the stars and the possible central black
hole\\ 
\vspace{-0.6cm}
\item the stellar mass-to-light-ratio is constant throughout the relevant range ($
< 2''$)\\
\vspace{-0.6cm}
\item the stars are in spherical symmetry.\\
\vspace{-0.6cm}
\end{enumerate}
Here, we construct a model with a Multi-Gaussian-expansion both for the
stellar photometry and for the gas distribution. The gas surface brightness is
given by the HST [FeII] narrow band image \citep{marconi00} and fitted
by an MGE model.\\ 
Following \cite{tremaine94} in the spherical case the solution of Jeans
equation for the projected rms velocity reduces to 
$$ (v^2+\sigma^2)_{p}(R) = \frac{2 G}{\Upsilon \Sigma_g(R)} \int_R^{\infty}
\frac{\rho_g(r)M_{tot}(r)}{r^2} (r^2-R^2)^{1/2} dr, $$
where $\Upsilon$ is the stellar mass-to-light ratio, $\Sigma_g(R)$ is the
surface brightness of the gas, $\rho_g(r)$ is the density of the gas and
$M_{tot}(r)$ is the mass of the stellar body and the central dark object.\\
Again, the stellar mass-to-light ratio is fixed at a value
of $\Upsilon= 0.72 \rm{M}_{\odot} / \rm{L}_{\odot}$, that we derived
in section \ref{stellar_mass}. 
The best fitting black hole mass that we find with the spherical Jeans model
is $(1.0\pm 0.5) \times 10^8 M_{\odot}$. 
The comparison between the measured
and the modelled rms velocity ($\sqrt{v^2+\sigma^2}$) is shown in Figure
\ref{spherical_jeans_gas}. 

\section{DISCUSSION}

The availability of adaptive optics instrumentation opens up new realms of ground
based observations. Drawing on kinematic data with unprecedented spatial resolution we present a
dynamical model of the black hole in the nucleus of Centaurus~A.
The spatial resolution that we reach in our 
H-band spectra is $0\farcs11$ (= 1.8 pc at a distance of 3.5 Mpc) that is a factor of 3 to 4 higher than from the
previous ground based observations with ISAAC by \cite{marconi01}. At large
radii ($>0\farcs3$) our data are in agreement with their gas kinematical data,
but our high resolution data reveals the steep gradient of
the rotation curve in much more detail (see e.g. Figure \ref{rot}). The
rotation curves show a smooth behavior and the slit position that shows the
fastest rotation (P.A.=$-44\fdg5$) is nearly perpendicular to the jet
direction (P.A.$\sim 51\degr$). This suggests the picture of gas orbiting in
a flattened geometry (presumably a disk) around the central black hole, with
the jet pointed along the disk angular momentum vector.\\ 
 
\subsection{Black hole mass}

In this paper we present three different dynamical modeling approaches to describe the
gas kinematics in the central $\sim 2'' \simeq$ 33pc of
NGC5128. Two of the models are clearly conceptually inconsistent with the
data, either neglecting the dominant velocity dispersion of the gas (Model 1; cold disk), or
the disk-like geometry of the velocity field (Model 3; isotropic, spherical
Jeans). We have laid out these 
extreme models as limiting cases, which would result in a broad mass range, $2.7 \times 
10^{7}M_{\odot}$ $<$ M$_{\rm{bh}}$ $<$ $1.5 \times 10^{8}M_{\odot}$, and to
demonstrate how important the gas physics is, even in light of high resolution data.
Our physically most plausible model is a hot disk model at an inclination angle of
$45\degr$; the corresponding black hole mass is M$_{\rm{bh}} =
6.1^{+0.6}_{-0.8} \times 10^{7}M_{\odot}$ with a line-of-nodes at $\zeta =
-27\degr \pm 4\degr$. 
We consider this model the best-fitting since it accounts for the high
velocity dispersion of the gas combined with a disk-like gas structure, as
suggested by the smooth rotation curves. The assumption of a
geometrically thin disk is not physically motivated but was chosen to
eliminate line-of-sight integrations through an underconstrained 3-dimensional
gas distribution.\\ 

Our disk model meshes well with other constraints on the central geometry of
NGC~5128: e.g. \cite{tingay98} derived a jet position
angle of $\sim 51\degr$. If we assume that the accretion disk and surrounding
gas disk is at right angles, we would expect its line-of-nodes to be at $\sim
-39\degr$. Moreover, from 
the direction of the jet and counter-jet they derived a value for the jet
inclination of $50\degr < i < 80 \degr$ with respect to the line-of-sight. 
Our best-fitting inclination angle of $45\degr$ is close to their lower value,
but our long-slit data do not tightly constrain 
the inclination angle. This was also the main source of uncertainty
for the black hole mass in the previous study by \cite{marconi01}. If we were
to consider a gas disk inclination to be as face-on as 25$\degr$ (inconsistent with
the jet inclination), the black hole mass would increase to $\sim1.2 \times
10^{8}M_{\odot}$ in the hot disk case. 

The unsettled question of inclination will presumably be
solved with the analysis of integral field spectroscopy data taken with
SINFONI at the VLT. The full 2-D velocity field data provide an excellent
means of modelling black hole masses from gas as well as stellar kinematics.\\

Our best-fit value for the line-of-nodes is only $\sim 12\degr$
away from the ``expected'' value of $\sim -39\degr$, much closer than
the value derived by \cite{marconi01}, $-10\degr > \zeta >
-14\degr$. Nonetheless, this confirms that the kinematical line-of-nodes does
not coincide with the projected major axis of the gas disk (P.A.$\simeq33\degr$)
seen in Pa$\alpha$ and [FeII] with NICMOS \citep{schreier98, marconi00}.\\ 
The value that we derived for the black hole mass from the best-fitting
hot disk model ($6.1^{+0.6}_{-0.8} \times 10^{7}M_{\odot}$) 
is significantly lower than 
the values both from stellar kinematics presented by \cite{silge05}
(M$_{\rm{bh}} = 1.8^{+0.4}_{-0.4} \times 10^{8}M_{\odot}$ for i=45$\degr$) and
the previous gas kinematical study of \cite{marconi01} (M$_{\rm{bh}} =
2.0^{+3.0}_{-1.4} \times 10^{8}M_{\odot}$) obtained at 3-4 times lower
resolution.  This decrease in black hole mass estimate with higher resolution data is
in line with the decrease in derived black hole masses when HST data became
available. 
Our limiting case model of a spherical gas distribution modeled through Jeans
equation gives a black hole mass of M$_{\rm{bh}} = (1.0 \pm 0.5) \times
10^{8}M_{\odot}$ which agrees both with \cite{silge05} and
\cite{marconi01}. 
However, in their gas dynamical model
\cite{marconi01} assumed a cold thin disk and did not account for the pressure
support of the gas; assuming a cold disk model, we find the best-fitting black
hole mass to be M$_{\rm{bh}} = (4.0^{+0.2}_{-1.0}) \times 10^{7}M_{\odot}$, which is
almost a factor of 7 lower than their value. \\

\subsection{Relation of black hole mass versus galaxy properties}

This confirmation of a fairly high black hole mass compared to a fairly low
stellar velocity dispersion of $\sigma_{*}=138$ km s$^{-1}$ \citep{silge05}
quantitatively reduces but qualitatively confirms the offset of Centaurus~A
from the M$_{\rm{bh}}$-$\sigma$ 
relation \citep{ferrarese00, gebhardt00}. The black hole mass
predicted by this relation would be around $3 \times 10^{7}M_{\odot}$ and
our best-fitting black hole mass lies a factor of $\sim 2$ above this. This offset is close to the
observed scatter of the M$_{\rm{bh}}$-$\sigma$ which is a factor of 1.5 (0.2dex)
\citep{ferrarese00, gebhardt00}.
In the case of the lowest black hole mass supported by our cold
disk model (M$_{\rm{bh}} = (4.0^{+0.2}_{-1.0}) \times 10^{7}M_{\odot}$) the black hole
falls nicely onto this relation.\\
To compare the black hole mass to the bulge mass of
Centaurus~A, we applied a spherical Jeans model to the whole galaxy
as seen in K-band, using the stellar kinematics (v$_*$ and $\sigma_*$) of
\cite{silge05} out to $\sim100\arcsec$ and derived a total mass of the spheroid of M$_{\rm{sph}} =
(2.5 \pm 1) \times 10^{11}M_{\odot}$. This mass is in agreement with the value M$_{\rm{sph}} =
(4 \pm 1) \times 10^{11}M_{\odot}$ derived by \cite{hui93} and \cite{mathieu99}.
Given this spheroid mass, Cen~A lies below the M$_{\rm{bh}}$-M$_{\rm{Bulge}}$
relation \citep[e.g.][]{haering04} (i.e. it has a fairly low black hole mass compared
to its bulge mass) but is not a striking outlier in this
relation. 
Taken together, it seems that Cen~A foremost has a very high
M$_{\rm{sph}}$/$\sigma_{*}$ ratio among ellipticals (which implies a very
low concentration), rather than being an outlier in the relations to
M$_{\rm{bh}}$. This low concentration may be explained by the fact that Cen~A is
known to be a z$\simeq$0 merger \citep{israel98}.\\

This paper demonstrates that
near-IR adaptive optics instrumentation provide excellent data
and make it possible to explore the central regions of dust enshrouded
galaxies. With a spatial resolution of $\sim 0\farcs06$ in K-band and $\sim
0\farcs11$ in H-band we are now able to resolve the radius of influence of
black holes even in more distant galaxies from the ground. 

\acknowledgments
This work is based on observations collected at the European Southern
Observatory, Paranal, Chile, ESO Program 72.B.0294A. We thank the Paranal
Observatory Team for the support during the observations. N. Neumayer thanks
the group of T. de Zeeuw at Leiden Observatory for their hospitality. MC
acknowledges support from a VENI award 639.041.203 awarded by the Netherlands
Organization for Scientific Research (NWO). This
publication makes use of data products from the Two Micron All Sky Survey,
which is a joint project of the University of Massachusetts and the Infrared
Processing and Analysis Center/California Institute of Technology, funded by
the National Aeronautics and Space Administration and the National Science
Foundation.

\end{document}